\documentstyle[aps,prl]{revtex}
\begin{document}
\twocolumn[\hsize\textwidth\columnwidth\hsize\csname %
@twocolumnfalse\endcsname
\draft
\title{Numerical Renormalization-Group Study of Particle-Hole Symmetry Breaking
\\
in Two-Channel Kondo Problem :\\
Effect of Repulsion between Conduction Electrons and Potential Scattering}
\author{H. Kusunose, K. Miyake, Y. Shimizu$^{*}$, and O. Sakai$^{**}$}
\address{Department of Material Physics, Faculty of Engineering Science,
Osaka University, \\ Toyonaka 560, Japan}
\address{$^{*}$Department of Applied Physics, Tohoku University, Sendai 980,
Japan}
\address{$^{**}$Department of Physics, Tohoku University, Sendai 980, Japan}
\maketitle
\begin{abstract}
Particle-hole symmetry breaking perturbation in two-channel pseudospin Kondo
problem is studied by the numerical renormalization-group method.
It is shown that the repulsion between conduction electrons at the impurity
site and the single particle potential are the relevant perturbations against
the conventional non-Fermi liquid fixed point.
Although the repulsion (potential) with realistic strength prevents the
overscreening of pseudospin, it induces in turn a {\it real spin}, which is
also overscreened again. Thus the {\it real spin} susceptibility becomes
anomalous contrary to the conventional two-channel Kondo problem.
\end{abstract}
\pacs{PACS numbers: 71.27.+a, 71.28.Mb, 74.70.Ad}
\vspace{2mm}
]
Multichannel Kondo problem has attracted much attentions in these years
because of its anomalous non-Fermi liquid behavior.
While the problem was originally discussed long ago as a generalized Kondo
effect of magnetic impurity with orbital degeneracy\cite{nozieres}, the
two-channel Kondo problem has recently been revived in a proposal of
quadrupolar Kondo effect as an origin of U-based heavy fermions\cite{cox}.
The two-level system interacting with conduction electrons was also
recognized as a candidate for realization of two-channel Kondo
model\cite{muramatsu}.
The latter system has attracted much interest not only because it offers a
model explaining anomalous transport properties of glassy
metals\cite{blume,vladar} but also it is expected to give a canonical model
of strong coupling electron-phonon systems\cite{matsuura,anderson,ayache}.

Although the two-channel Kondo problem has been fully solved by a variety of
methods\cite{cragg,pang,sakai,andrei,tsvelik,emery,coleman}, there seems
still to remain for us to clarify a reality of the model itself.
As for relation with experiments, the pseudospin two-channel Kondo model
gives anomalous behavior the susceptibility of the pseudospin (i.e., that of
charge polarization), while it is the real spin susceptibility that exhibits
non-Fermi liquid behaviors in some of the above systems.
The latter point may be related to an appearance of localized real spin.
For instance, a repulsion between conduction electrons at the impurity site,
which was neglected in the above pseudospin models\cite{mspin}, is expected
to prevent the overscreening\cite{miyake} and to induce a real spin.

The purpose of this paper is to examine the effect of such a repulsion on
the two-channel Kondo model by numerical renormalization-group (NRG)
method\cite{wilson,krishna}.
It is shown that the fixed point Hamiltonian is described not only by the
conventional exchange coupling $J^*$ but also an impurity potential $V^*$,
to which the repulsive interaction $\tilde{U}$ is renormalized.
Namely, the single particle potential $\tilde{V}$ is also a relevant
perturbation.
The competition between the exchange coupling and the repulsion or the
impurity potential induces the degrees of freedom of channel (i.e., real spin)
and leads to the pseudospin singlet ground state for realistic strength
of $\tilde{U}$ or $\tilde{V}$.
It is the particle-hole symmetry breaking that causes such competition.
The overscreening of so induced real spin makes again the real spin
susceptibility anomalous contrary to the conventional pseudospin two-channel
Kondo problem without $\tilde{U}$ and $\tilde{V}$.

We begin with the model Hamiltonian for the Wilson NRG calculation as follows:
\begin{eqnarray}
\label{e5}
&&\frac{H_N}{D}=\Lambda^{(N-1)/2}\{\sum_{m\sigma}\sum_{n=0}^{N-1}
\Lambda^{-n/2}\xi_n(f^{\dagger}_{n,m\sigma}f_{n+1,m\sigma}+{\rm h.c.})
\nonumber \\
&&\mbox{\hspace{5.5cm}}+H_{\rm int}\},
\end{eqnarray}
where
\begin{eqnarray}
\label{e6}
&&H_{\rm int}=J\sum_{m\sigma\sigma'}f^{\dagger}_{0,m\sigma'}
\mbox{\boldmath $\sigma$}_{\sigma'\sigma}f_{0,m\sigma}\cdot\mbox{\boldmath
$\tau$}
+V\sum_{m\sigma}f^{\dagger}_{0,m\sigma}f_{0,m\sigma} \nonumber \\
&&\mbox{\hspace{1cm}}+\frac{U}{2}\sum_{mm'\sigma\sigma'}^{m\sigma\neq
m'\sigma'}
f^{\dagger}_{0,m\sigma}f_{0,m\sigma}f^{\dagger}_{m'\sigma'}f_{0,m'\sigma'},
\end{eqnarray}
where the indices $m$ and $\sigma$ denote a label indicating channel $1$, $2$
and pseudospin $\uparrow$, $\downarrow$, respectively, $\mbox{\boldmath
$\sigma$}$ is the
Pauli matrix vector, and $\mbox{\boldmath $\tau$}$ is twice the operator of the
impurity
pseudospin of $1/2$.
The pseudospin stands for the charge polarization at the impurity site and
two channels for two components of ``real" spin of conduction electron.
Here we have defined
\begin{eqnarray}\label{e8}
&&D\equiv\frac{1+\Lambda^{-1}}{2}\tilde{D},\mbox{\hspace{10mm}}
J\equiv\frac{1}{1+\Lambda^{-1}}\tilde{J},\nonumber \\
&&U\equiv\frac{8}{1+\Lambda^{-1}}\tilde{U},\mbox{\hspace{10mm}}
V\equiv\frac{4}{1+\Lambda^{-1}}\tilde{V},
\end{eqnarray}
where $2\tilde{D}$ denotes the bandwidth of conduction electrons, $\tilde{J}$
the exchange interaction between conduction electrons and impurity pseudospin,
$\tilde{V}$ the potential at impurity site, and $\tilde{U}$ the repulsion
between conduction electrons at impurity site\cite{aniso}.
Hereafter we set $D=1$, i.e., the energy levels are scaled by $D$ and ignore
$\Lambda$-dependence in $\xi_n$, i.e., $\xi_n = 1$, because
$\xi_n\rightarrow 1$ for large $n$.

The conserved quantities of the Hamiltonian $H_N$, (\ref{e5}), are the total
number of conduction electrons $Q$, the real spin $j$ and the total
pseudospin $S$, defined as follows:
\begin{eqnarray}
\label{e10}
&&Q_N=\sum_{m}\sum_{n=0}^{N}\sum_{\sigma}(f^{\dagger}_{n,m\sigma}f_{n,m\sigma}-1/2),\\
\label{e11}
&&{\bf j}_N=\frac{1}{2}\sum_{\sigma}\sum_{n=0}^{N}\sum_{mm'}
f^{\dagger}_{n,m'\sigma}\mbox{\boldmath$\sigma$}_{m'
m}f_{n,m\sigma}\equiv\sum_{\sigma}
{\bf j}_{\sigma}^{N},\\
&&{\bf S}_N=\frac{1}{2}\left[\sum_{m}\sum_{n=0}^{N}\sum_{\sigma\sigma'}
f^{\dagger}_{n,m\sigma'}\mbox{\boldmath$\sigma$}_{\sigma'\sigma}f_{n,m\sigma}+\mbox{\boldmath$\tau$}\right]
\nonumber \\
&&\mbox{\hspace{5mm}}\equiv\sum_{m}{\bf s}_{m}^{N}+{\bf t}.
\end{eqnarray}
Since both the repulsion $U$ and the potential $V$ breaks the particle-hole
symmetry unless $3U/2+V=0$, the degenerate eigenstates denoted by $\pm Q$
are split in general.
In our calculations, we have used $\Lambda =3$ and retained low lying energy
states up to 300 states at each step as bases for constructing new quadruple
states.

First we have investigated the case $U=V=0$ and verified that the same energy
levels as the work of Pang and Cox\cite{pang} are reproduced.
Next we have investigated the case $U\neq0$.
The flow diagram of level of low lying states for $J=2.0$ and $U=1.6$ is shown
in Fig. \ref{f3}.
The solid (dotted) lines are for even (odd) iterations.
Each level is labelled by $(Q,j,S)$.
The ground state of the fixed point is pseudospin doublet ($S=1/2$), which is
expected for the case where the exchange coupling is stronger than the
repulsion.
In Fig. \ref{f4} the flow diagram for $J=1.0$ and $U=2.0$ is shown.
The ground state is now pseudospin singlet ($S=0$), because the repulsion $U$,
larger than exchange coupling $J$ and the hopping $D=1$, prohibit the
overscreening.
It is noted that the ground state is still degenerate due to degrees of
freedom of channel, i.e., $j=1/2$.
It is remarked that the position of energy levels at the fixed point in Fig.
\ref{f3} and Fig. \ref{f4} exactly coincide with each other while the nature
of ground state is different.
As discussed later, the quantum numbers specifying each levels in Fig.
\ref{f3} are in a certain relation with those in Fig. \ref{f4}.

The nature of ground states for various coupling constants $J$, $U$ and
$V=0$ are shown in Fig. \ref{f7}.
The closed circles stand for the ground state with $S=0$, while the open
circles $S=1/2$. The line dividing two types of ground state is drawn by
estimating the coupling constants which give the same energies of these
two types of ground state.
It is noted that the boundary line flatten as $J\rightarrow0$ and has a
constant slope for $J {\lower -0.3ex \hbox{$>$} \kern -0.75em \lower 0.7ex
\hbox{$\sim$}} 1$.
We can understand this result as follows.
The energy gains for overscreening formation at the impurity ($n=0$) site
are both due to the exchange energy $J$ and the kinetic energy associated
with the transfer $D$ between 0- and 1-site, while the energy loss arises
through the repulsion $U$ between overscreened conduction electrons.
On the other hand, for singlet formation, the energy gain is due to avoiding
the repulsion $U$ while the energy loss is due to the exchange coupling $J$
and the kinetic energy $D$.
Consequently, the boundary line is roughly determined by the condition
$U\sim \max(J,D)$.
Namely, for $\tilde{U}>\tilde{J}/8$ and $\tilde{D}/4$, the ground state
becomes a pseudospin singlet.
It is noted that the ground state is expected to belong to that of $S=0$ for
a realistic value of $\tilde{U}$ and $\tilde{J}$.

Now we discuss about properties of the fixed point.
The fixed point Hamiltonian $H^*$ is described as
\begin{eqnarray}
&&H^* =\sum_{n=0}^{\infty}\Lambda^{-n/2}(f^+_{n+1}f_n+{\rm h.c.})
+ H_{\rm int}^*, \nonumber \\
\label{fixint}
&&\mbox{\hspace{5mm}}H_{\rm int}^*
=4J^*\left({\bf s}_1^0+{\bf s}_2^0\right)\cdot{\bf t}
+\alpha^*\left({\bf s}_1^0+{\bf s}_2^0\right)^2
\nonumber \\
&&\mbox{\hspace{2.5cm}}+\beta^*\left({\bf j}_\uparrow^0
+{\bf j}_\downarrow^0\right)^2+V^* Q_0+\epsilon,
\end{eqnarray}
where $H_{\rm int}^*$ has the same symmetry as $H_{\rm local}$, in (\ref{e5}),
the effective couplings $J^*$, $\alpha^*$, $\beta^*$, and $V^*$ may depend on
the initial couplings in general, and $\epsilon$ is a constant energy shift.
If we set $J^*=J$, $\alpha^*=\beta^*=-U$, $V^*=3U/2+V$, and $\epsilon=3U+2V$,
$H_{\rm int}^*$ becomes equivalent to $H_{\rm local}$ in (\ref{e5}).

Since the energy of low lying excited states at the fixed point are mainly
determined by $H_{\rm int}^*$, we can determine the parameters
$J^*\sim\epsilon$ in (\ref{fixint}) so as to reproduce the low lying energy
levels at $N=39$.
The results for the initial parameters, ($J$,$U$,$V$), (a) (0.5,0.0,0.0),
(b) (2.0,1.6,0.0), (c) (1.0,2.0,0.0) and (d) (0.2,0.4,-0.6) are shown in
Table \ref{fv}.
It is noted that the effective exchange coupling, $J^*$ is independent of the
initial coupling $U$ (including $U=0$), and $\alpha^*$ and $\beta^*$ are
always zero.
In the case (d), there is the particle-hole symmetry so that the fixed point
is the same as the case (a) where $U=V=0$.
The character of the fixed point is determined mainly by the effective
impurity potential $V^*$ which depends on the initial couplings $J$, $U$ and
$V$, i.e. $V^*=f(J,U,V)$.
Consequently, the effective interaction at the fixed point can be written as
\begin{equation}
\label{fint}
H_{\rm int}^* = J^*({\bf s}_1^0+{\bf s}_2^0)\cdot{\bf t}+V^*Q_0,
\mbox{\hspace{1cm}} (J^*=0.20).
\end{equation}
$J$, $U$-dependence of $V^*$ with $V=0$ is shown in Fig. \ref{fd}.
It is noted that $V^*$ increases (decreases) as $U$ ($J$) increases.
{}From this effective interaction, the ``flow lines" for scaling in parameter
space are obtained from $f(J,U,V=0)= {\rm constant}$.
Especially, for $V^*=J^*=0.20$, the ``flow line" becomes equivalent to the
boundary line shown in Fig. \ref{f7}, because the first excitation energy is
zero for these couplings.

In order to investigate $\omega$ and $T$ dependence of susceptibility, let
$(Q,j,S)$ be $(Q_D,j_D,S_D)$ for $V^*=V^*_D<0.20$, where the ground state
is pseudospin doublet ($S=1/2$), and $(Q_S,j_S,S_S)$ for $V^*=V^*_S>0.20$
where the ground state is pseudospin singlet ($S=0$).
For each low lying excited states, we can find the relations, $Q_S=-Q_D-1$,
$j_S=S_D$, and $S_S=j_D$.
If we set $V^*_S=2J^*-V^*_D$, the low lying excited energies at the fixed
point for each parameters are the same as easily seen by means of the
effective interaction (\ref{fint}).
A prime example is the relation between Fig. \ref{f3} and Fig. \ref{f4} as
mentioned above.
According to this example, the coincidence of energy levels occurs after 20
iterations.
{}From this coincidence, it is expected that $\omega$ and
$T$($\omega/D,T/D<\Lambda^{-20/2}$) dependence of the susceptibility of
{\it real spin} (channel) for $V^*_D$ coincide with those of the pseudospin
for $V^*_S=2J^*-V^*_D$ and vice versa.
This is a new aspect of two-channel Kondo problem which has not been
recognized so long as the conventional model without the repulsion and the
potential scattering ($U=V=0$) had been investigated.

This remarkable aspect can be seen more vividly by investigating the spectral
weight of the dynamical susceptibilities for real spin of conduction
electrons at the impurity site, $\chi_j''(\omega)$, and for the impurity
pseudospin, $\chi_t''(\omega)$.
They are calculated by the method of ref. \cite{sakai} as shown in Figs.
\ref{ff1} and \ref{ff2}.
It is noted that, in the presence of the repulsion $U$, $\chi_j''(\omega)$
shows the non-Fermi liquid behavior with
$\lim_{\omega\rightarrow 0}\chi_j''(\omega)$ being finite, while without
the repulsion it shows Fermi liquid one with $\chi_j''(0)=0$.
However, if we set $3U/2+V=0$, the similar calculations show that
$\chi_j''(0)=0$.
Namely, it is the particle-hole symmetry breaking that gives
$\chi_j''(\omega)$'s the non-Fermi liquid behaviors.
The potential $V^*$ shifts the number of conduction electrons at the impurity
site from one in each channel, though the exchange works to hold the
overscreening formation.
This competition induces the degrees of freedom of channel (i.e., real spin).
It is also overscreened again by conduction electrons with two channels, i.e.,
pseudospin degrees of freedom.
Thus the real spin susceptibility becomes anomalous due to the potential $V^*$
which breaks the particle-hole symmetry.
The case $V^*<0$ is understood as $V^*>0$ by particle-hole transformation.
Since the degrees of freedom of pseudospin, however, are not perfectly
vanished, pseudospin susceptibility is anomalous for any strength of the
repulsion including $U=0$.

In summary, the low lying excited states at the fixed point of pseudospin
two-channel Kondo model with the particle-hole symmetry breaking perturbations
are described not only by the exchange $J^*$ but also the impurity potential
$V^*$. For $|V^*|>0.20$, i.e., $\tilde{U}>\tilde{J}/8$ and $\tilde{D}/4$,
realistic values, a pseudospin singlet ground state is realized in contrast
with the pseudospin doublet ground state which is realized in the conventional
two-channel Kondo problem.
The spectral weight of the dynamical susceptibility of the real spin shows
the non-Fermi liquid behavior because of the overscreening of the real spin.
Thus, it is expected that the {\it magnetic} non-Fermi liquid behaviors
observed in some compounds are understood by the particle-hole symmetry
breaking perturbation, which induces the degrees of freedom of the real spin.

We have much benefited from conversations with Y. Kuramoto.
Two of the authors (H. K. and K. M.) express their appreciation to
P. Nozi\`eres for a critical correspondence at very early stage of this work
and to K. Ueda for clarifying discussions on the results.
This work is supported in part by a Grant-in-Aid for Scientific Research on
Priority Areas: ``Physics of Strongly Correlated Conducting Systems"
(06244104) from the Ministry of Education, Science and Culture.

\begin{figure}
\caption{The flow diagram for $J=2.0$, $U=1.6$ and $V=0$. Each level is
labelled by $(Q,j,S)$, where $Q$, $j$, and $S$ are the total number of
electrons, the real spin, and the total pseudospin, respectively. The solid
(dotted) lines are for even (odd) iterations. The ground state is a pseudospin
doublet.}
\label{f3}
\end{figure}

\begin{figure}
\caption{The flow diagram for $J=1.0$, $U=2.0$ and $V=0$. The ground state is
a pseudospin singlet. The degeneracy of the ground state is due to degrees of
freedom of channel, i.e., $j=1/2$.}
\label{f4}
\end{figure}

\begin{figure}
\caption{The nature of ground states for various sets of coupling constants
of $J$, $U$ and $V=0$ in the unit $D$. The open circles are for the ground
state with pseudospin doublet $S=1/2$, while the closed circles the ground
state with pseudospin singlet $S=0$. The line dividing two phases of, $S=0$
and $S=1/2$, is drawn by inspection.}
\label{f7}
\end{figure}

\begin{figure}
\caption{$J$, $U$-dependence of the effective impurity potential, $V^*$, with
$V=0$ at the fixed point.}
\label{fd}
\end{figure}

\begin{figure}
\caption{$\omega$ dependence of imaginary part of local real spin
susceptibility $\chi_j''(\omega)$.}
\label{ff1}
\end{figure}

\begin{figure}
\caption{$\omega$ dependence of imaginary part of impurity pseudospin
susceptibility $\chi_t''(\omega)$.}
\label{ff2}
\end{figure}

\begin{table}
\begin{center}
\begin{tabular}{c|c|c|c|c|c|c}
 &($J$,$U$,$V$) & $J^*$ & $\alpha^*$ & $\beta^*$ & $V^*$ & $\epsilon$\\ \hline
(a)& (0.5,0.0,0.0) & 0.20 & 0 & 0 & 0 & 0.80 \\ \hline
(b)& (2.0,1.6,0.0) & 0.20 & 0 & 0 & 0.12 & 0.80 \\ \hline
(c)& (1.0,2.0,0.0)& 0.20 & 0 & 0 & 0.28 & 0.88 \\ \hline
(d)& (0.2,0.4,-0.6)& 0.20 & 0 & 0 & 0 & 0.80
\end{tabular}
\end{center}
\caption{Effective couplings, $J^*\sim\epsilon$, which make a reproduction of
the energy levels at $N=39$ for the initial parameters, ($J$,$U$,$V$),
(a) (0.5,0.0,0.0), (b) (2.0,1.6,0.0), (c) (1.0,2.0,0.0), and
(d) (0.2,0.4,-0.6).}
\label{fv}
\end{table}

\begin{references}
\bibitem{mspin}Such a repulsion has already been taken into account at the
level of the ``Anderson" model for the magnetic two-channel Kondo problem.
\bibitem{aniso}Since the Coulomb repulsion works most effectively between
conduction electrons with the same polarization, the repulsion is anisotropic
in pseudospin space.
However, the anisotropic interaction is shown to flow into an isotropic one
by renormalization-evolution as will be discussed elsewhere; so that we
confine ourselves to discussing the case of isotropic repulsive interaction
in (2).
\end{references}
\end{document}